\documentclass[aps,prl,twocolumn,superscriptaddress,groupedaddress,footinbib]{revtex4-1} 

\usepackage{longtable}
\usepackage{morefloats}
\usepackage[dvips]{graphicx}
\usepackage{color}
\usepackage{epsfig,graphicx,amsfonts,amsbsy}
\usepackage{amsmath,amsfonts,amsthm,amssymb}
\usepackage{appendix}
\usepackage{makeidx}
\usepackage{url}
\usepackage{verbatim}
\usepackage[bookmarksnumbered,pdfpagelabels=true,plainpages=false,colorlinks=true,linkcolor=blue,citecolor=red,urlcolor=blue]{hyperref}
\usepackage[rightcaption]{sidecap}
\usepackage{array}
\usepackage{multirow}
\usepackage{tabularx}
\usepackage{braket} 
\usepackage{dsfont}
\usepackage{bbold}
\usepackage{etoolbox}

\newcommand{\bs}[1]{\boldsymbol{#1}}
\newcommand{\ssf}[1]{\mathsf{#1}}

\def\half{{\tfrac{1}{2}}} 

\def\Br{{\bs{r}}}
\def\Bk{{\bs{k}}}
\def\Be{{ \bs{e} }}
\def\BR{{\bs{R}}}
\def\BS{{\bs{S}}}
\def\BD{{\bs{D}}}
\def\BPsi{{\bs{\Psi}}}

\def\SS{{\ssf{S}}}

\def\zhat{{ \bs{\hat{z}} }}

\def\muB{{ \mu_{\mbox{\tiny B}} }}

\def\CalO{{\mathcal{O}}}
\def\CalH{{\mathcal{H}}}
\def\CalE{{\mathcal{E}}}
\def\CalN{{\mathcal{N}}}

\def\Hsw{{H_{\mbox{\tiny SW}}}}

\def\Tr{{\mbox{Tr}}}

\begin{document}

\title{Topological Magnons and Edge States in Antiferromagnetic Skyrmion Crystals}

\author{Sebasti{\'a}n A. D{\'i}az}
\author{Jelena Klinovaja}
\author{Daniel Loss}

\affiliation{Department of Physics, University of Basel, Klingelbergstrasse 82, CH-4056 Basel, Switzerland}

\date{\today}
	
\begin{abstract}
Antiferromagnetic skyrmion crystals are magnetic phases predicted to exist in antiferromagnets with Dzyaloshinskii-Moriya interactions. Their spatially periodic noncollinear magnetic texture gives rise to topological bulk magnon bands characterized by nonzero Chern numbers. We find topologically-protected chiral magnonic edge states over a wide range of magnetic fields and Dzyaloshinskii-Moriya interaction values. Moreover, and of particular importance for experimental realizations, edge states appear at the lowest possible energies, namely, within the first bulk magnon gap. Thus, antiferromagnetic skyrmion crystals show great promise as novel platforms for topological magnonics.
\end{abstract}

\maketitle

%%%%%%%%%%%%%%%%%%%%%%%%%%%%%%%%%%%%%%%%%%%%%%%%%%%%%%%%%%%%%%%%%%%%%%
%%INTRODUCTION
%%%%%%%%%%%%%%%%%%%%%%%%%%%%%%%%%%%%%%%%%%%%%%%%%%%%%%%%%%%%%%%%%%%%%%

Interest in antiferromagnets has been recently renewed in part due to progress in the theoretical understanding of their interaction with electric currents \cite{Nunez2006,Haney2007,Shick2010,Zelezny2014} that culminated in the experimental demonstration of the electrical switching of CuMnAs \cite{Wadley2016} and the consequent prospect of future applications. Apart from being more ubiquitous in Nature than ferromagnets, antiferromagnets enjoy attractive properties: lack of stray fields, insensitiveness to external magnetic field perturbations, and ultrafast dynamics in the THz range \cite{Keffer1952,Bossini2016,Roy2016,Olejnik2018}. Moreover, spintronics research on antiferromagnets has led to the study of spin currents carried by magnetic excitations \cite{Meier2003,Chumak2015,Nakata2017a,Baltz2018,Jungfleisch2018,Lebrun2018}. However, losses due to Joule heating hinder the design of  low-power spintronic devices, making insulating antiferromagnets without itinerant electrons most preferable candidate materials \cite{Trauzettel2008}.

Besides low dissipation, another sought-after property is the ability to reliably control magnon spin currents. Inspired by insights from topological matter research \cite{Hasan2010}, suitable candidates where this could be achieved are the recently predicted topological magnonic insulators \cite{Shindou2013,Zhang2013,Mook2014,Owerre2016,Owerre2017,Mook2017,Nakata2017b,Nakata2017c,Seshadri2018,Li2018,McClarty2018}. These spatially periodic magnetic systems possess topological bulk magnon bands and hence could host topologically-protected chiral or helical edge states, i.e., stable unidirectional channels for magnon spin currents localized at the edges of the sample. One route to realize them, which allows high tunability, exploits the emergent gauge fields of magnetic textures \cite{Nagaosa2012a,Nagaosa2012b,vanHoogdalem2013} induced by Dzyaloshinskii-Moriya interactions (DMI), as shown in particular for ferromagnetic \cite{vanHoogdalem2013,Roldan-Molina2016} and ferrimagnetic \cite{Kim2018} skyrmion crystals (SkX).

\begin{figure}[t!]
\centering
\includegraphics[width=\columnwidth]{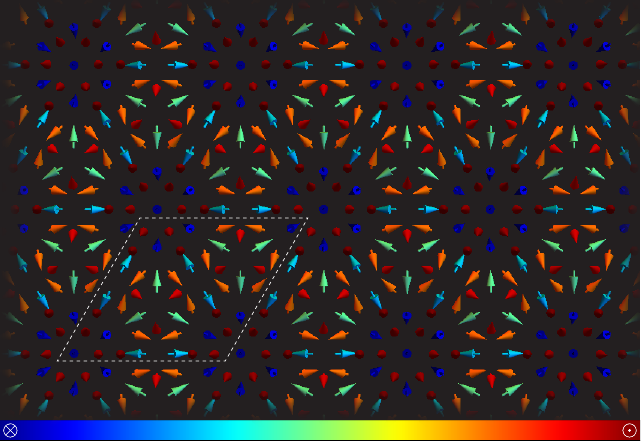}
\caption{Classical ground state texture of the AFM-SkX for the model parameters $D/J = 0.5$ and $g\muB B/JS = 2.5$. The magnetic unit cell, comprised of 7$\times$7 spins enclosed by the white dashed-line loop, has $Q = -1$ topological charge. The color code indicates the out-of-plane component of the spins.}
\label{fig:BulkTexture}
\end{figure}

Motivated by the promising traits of antiferromagnets and the need for low-dissipation materials, in this letter we report our numerical study of the topology of the magnonic excitations supported by an electrically insulating antiferromagnetic skyrmion crystal (AFM-SkX) \cite{Rosales2015,Gobel2017}, depicted in Fig. \ref{fig:BulkTexture}. An important advantage of AFM-SkX's over previously considered antiferromagnets \cite{Zyuzin2016,Owerre2017c} is the possibility to obtain topologically-protected magnonic edge states at sufficiently low energies (where level broadening due to magnon-magnon interactions is minimal \cite{Mook2016b}). This results from the larger number of spins in the magnetic unit cell, thus accommodating more magnon bands at lower energies. Furthermore, we find topologically-protected chiral magnonic edge states within the first bulk magnon gap, at the lowest possible energies they can exist and where the noninteracting magnon approximation adopted in the following is expected to work best.  

%%%%%%%%%%%%%%%%%%%%%%%%%%%%%%%%%%%%%%%%%%%%%%%%%%%%%%%%%%%%%%%%%%%%%%
%%MODEL/FORMALISM
%%%%%%%%%%%%%%%%%%%%%%%%%%%%%%%%%%%%%%%%%%%%%%%%%%%%%%%%%%%%%%%%%%%%%%

{\it Model.} We employ the spin-lattice Hamiltonian
\begin{align}\label{eq:SpinH}\nonumber
H &= - \half \sum_{< \Br, \Br' >} \Big(
J_{\Br,\Br'} \BS_\Br\cdot\BS_{\Br'} + \BD_{\Br,\Br'}\cdot\BS_\Br\times\BS_{\Br'} \Big) \\
& \hspace{70pt}- g\muB B \sum_\Br \BS_\Br\cdot\zhat \,,
\end{align}
with $\BS_\Br$ a spin-$S$ operator at site $\Br$  of a triangular lattice with lattice constant $a$ located on the $xy$-plane. It takes into account nearest neighbor antiferromagnetic exchange interaction and interfacial DMI. For $\Br$ and $\Br'$ being nearest neighbors: $J_{\Br,\Br'} = - J$ and $\BD_{\Br,\Br'} = D \, \zhat \times (\Br - \Br')/|\Br - \Br'|$. The Hamiltonian also includes coupling to an external magnetic field $B\zhat$, with $g$ and $\muB$ being the gyromagnetic ratio and Bohr magneton, respectively.

\begin{figure}[t!]
\centering
\includegraphics[width=\columnwidth]{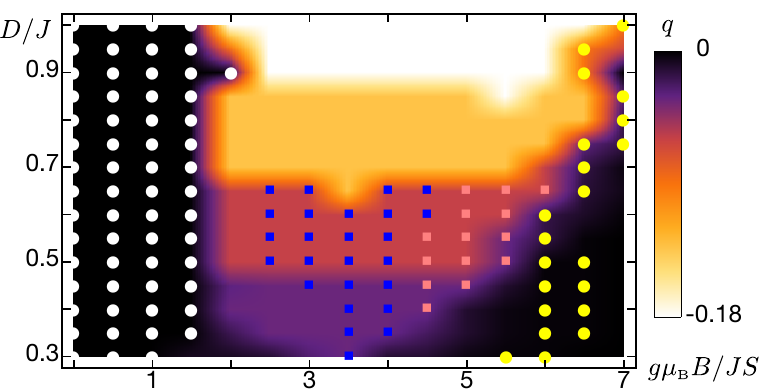}
\caption{Classical phase diagram at zero temperature. Among the magnetic textures are: helical (white dots), AFM-SkX (squares), and vortex crystal (yellow dots). Blue/pink squares represent AFM-SkX magnons with/without a global first bulk magnon gap. The color code indicates the topological charge density $q = 3Q/\CalN$, with $\CalN$ the total number of spins.}
\label{fig:PhDiag}
\end{figure}

Using Monte Carlo simulated annealing \cite{Evans2014} first and then time evolution according to a reduced version of the Landau-Lifshitz-Gilbert equations \cite{Berkov2005}, we build the classical phase diagram at zero temperature of this model as a function of $D/J$ and magnetic field, for systems with 28$\times$28 and 30$\times$30 spins (see Fig. \ref{fig:PhDiag}). The  static spin structure factor and the topological charge \cite{Rajaraman1989}, $Q$, are used to classify the magnetic phases. At low magnetic fields the helical phase is realized while at higher fields a crystal of magnetic vortices stabilizes. For intermediate magnetic fields the system favors the AFM-SkX, shown in Fig. \ref{fig:BulkTexture}, composed of three interpenetrating sublattices, each of which corresponds to a ferromagnetic SkX. While the AFM-SkX magnetic unit cell carries $Q = -1$ topological charge, each of the high-field magnetic vortices has $Q = -1/2$. We note that our result reproduces the special case $D/J = 0.5$ obtained before \cite{Rosales2015}.

In order to describe the quantum spin fluctuations about the AFM-SkX, we construct the Hamiltonian of free spin waves using Holstein-Primakoff (HP) bosons \cite{Holstein1940}. For the noncollinear texture of an AFM-SkX, it is convenient to choose the spin quantization axis along the direction of the classical ground state. This is accomplished by introducing an orthonormal basis at each site $\Br$, $\{ \Be_\Br^1, \Be_\Br^2, \Be_\Br^3 \}$, with $\Be_\Br^1 \times \Be_\Br^2 = \Be_\Br^3$ such that $\braket{\BS_\Br} = S\Be_\Br^3$. We can now rotate the spin operators according to $\BS_\Br = \Be_\Br^\alpha \SS_\Br^\alpha$. The HP transformation then reads $\SS_\Br^+ = (2S - a_\Br^\dag a_\Br)^{1/2} a_\Br$, $\SS_\Br^- = a_\Br^\dag(2S - a_\Br^\dag a_\Br)^{1/2}$, and $\SS_\Br^3 = S - a_\Br^\dag a_\Br$, where $\SS_\Br^\pm = \SS_\Br^1 \pm i \SS_\Br^2$.

When the magnetic unit cell of the classical ground state texture is comprised of more than one spin, as in the case of an AFM-SkX, the lattice site index $\Br$ can be partitioned as $\Br = \BR + \Br_j$. Here $\BR$ is a Bravais lattice vector and the $\Br_j$'s label the spins within the magnetic unit cell. Under this relabeling, the local basis vectors satisfy $\Be_{\BR j}^\alpha = \Be_j^\alpha$, the HP boson destruction operators are now written as $a_{\BR j}$ and are related to their lattice Fourier transforms through $a_{\BR j} = \tfrac{1}{\sqrt{N}}\sum_\Bk e^{i\Bk\cdot(\BR + \Br_j)}\, a_{\Bk j}$, with $N$ being the total number of magnetic unit cells. 

Following the standard procedure, the spin-lattice Hamiltonian is expanded as a series in $1/S$ \cite{Kittel1963}. Upon Fourier transformation the spin wave Hamiltonian, identified as the $\CalO (S)$ piece, is then given by
\begin{align}\label{eq:Hsw_reciprocal}
\Hsw = \half S \sum_{\Bk;i,j} \BPsi_{\Bk i}^\dag \CalH_{ij}(\Bk) \BPsi_{\Bk j} + \CalE_0 \,,
\end{align}
where $\CalE_0 = - \half N S\sum_i \Lambda_i$, $\BPsi_{\Bk i} = (a_{\Bk i} \,, a_{- \Bk i}^\dag)^T$, and
\begin{align}
\CalH_{ij}(\Bk) = 
\begin{pmatrix}
\Omega_{ij}(\Bk) & - \Delta_{ij}(\Bk) \\
- \Delta_{ij}^*(-\Bk) & \Omega^*_{ij}(-\Bk)
\end{pmatrix} \,.
\end{align}
Here, $\Delta_{ij}(\Bk) =  \half \big[ J_{ij}(\Bk) \, {\Be_i^+ \cdot \Be_j^+} + \BD_{ij}(\Bk) \cdot {\Be_i^+ \times \Be_j^+} \big]$, $\Omega_{ij}(\Bk) = \delta_{ij} \Lambda_i - \half \big[ J_{ij}(\Bk) \, {\Be_i^+ \cdot \Be_j^-} + \BD_{ij}(\Bk) \cdot {\Be_i^+ \times \Be_j^-} \big]$, $\Lambda_i = \sum_j \big[ J_{ij}(\Bk \!\!\! =\!\!\! 0) \, {\Be_i^3 \cdot \Be_j^3} + \BD_{ij}(\Bk \!\!\! =\!\!\! 0) \cdot {\Be_i^3 \times \Be_j^3} \big] + \frac{g\muB}{S}B \, \Be_i^3\cdot\zhat$, with $\Be_j^\pm = \Be_j^1 \pm i\,\Be_j^2$,  $J_{ij}(\Bk) = \sum_\BR J_{\BR + \Br_i , \Br_j } e^{- i\Bk \cdot ( \BR + \Br_i - \Br_j )}$ and similarly for $\BD_{ij}(\Bk)$.

It is possible to diagonalize this bosonic  Hamiltonian by a linear Bogoliubov transformation $(a_{\Bk i} \,, a_{- \Bk i}^\dag)^T = [T_\Bk]_{i,\lambda} (\alpha_{\Bk \lambda} \,, \alpha_{- \Bk \lambda}^\dag)^T$. To ensure that the new operators $\alpha_{\Bk \lambda}$ and $\alpha_{\Bk \lambda}^\dag$ satisfy the bosonic algebra, $T_\Bk$ must be paraunitary, i.e., $T_\Bk \, \sigma_z \, T_\Bk^\dag = \sigma_z$, where $\sigma_z$ acts on the subspace of $\alpha_{\Bk \lambda}$ and $\alpha_{- \Bk \lambda}^\dag$ \cite{Colpa1978}. The diagonalized spin wave Hamiltonian has the final form
\begin{align}
\Hsw = S \sum_{\Bk,\lambda} \CalE_{\Bk \lambda} \big( \alpha_{\Bk \lambda}^\dag \alpha_{\Bk \lambda} + \half \big) + \CalE_0 \,,
\end{align}
with $\CalE_{\Bk \lambda}$ the energy of the $\lambda$-th magnon band. The ground state of this Hamiltonian is defined by $\alpha_{\Bk \lambda} \ket{GS} = 0$, and $\ket{\Bk,\lambda} = \alpha_{\Bk \lambda}^\dag \ket{GS}$ represents a single magnon state. Note that the number of magnon bands is equal to the number of spins in the magnetic unit cell. In the following, we use the algorithm developed in Ref. \cite{Colpa1978} to calculate $T_\Bk$ numerically.

The Berry curvature of the $\lambda$-th magnon band is given by $B_\lambda(\Bk) = i\epsilon_{\mu\nu} \Tr \{ P_\lambda(\Bk) \, [\partial_{k_\nu} P_\lambda(\Bk)] \, \partial_{k_\mu} P_\lambda(\Bk) \}$ \cite{Shindou2013}, where $P_\lambda(\Bk) = T_\Bk \, \Pi_\lambda \, \sigma_z \, T_\Bk^\dag \, \sigma_z$ are projection operators, and $\Pi_\lambda$ is a diagonal matrix whose $\lambda$-th diagonal element is $+1$ and zero otherwise. Finally, the corresponding Chern number reads $C_\lambda = \int d\Bk \, B_\lambda(\Bk)/2\pi$, and the integration is over the first Brillouin zone (BZ).

\begin{figure}[t!]
\centering
\includegraphics[width=\columnwidth]{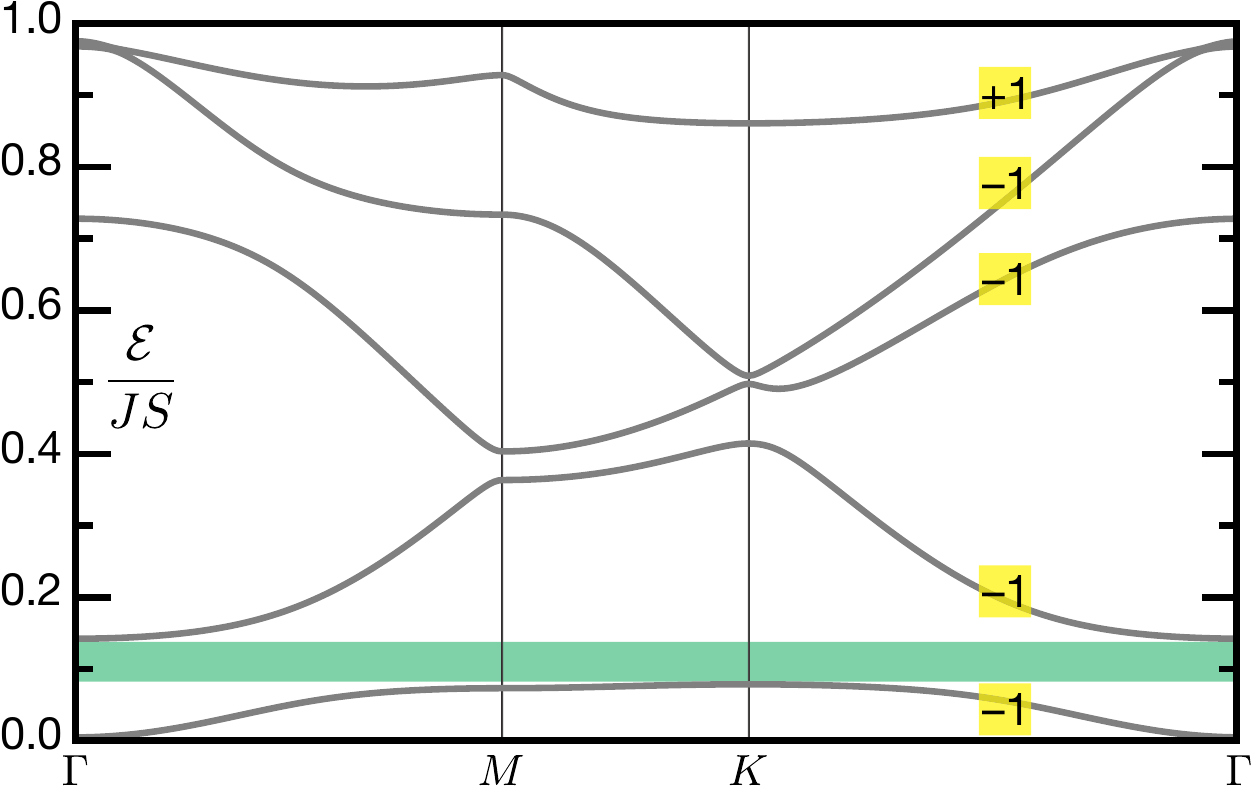}
\caption{Bulk magnon band structure of the AFM-SkX along the first BZ loop $\Gamma$-$M$-$K$-$\Gamma$ for the same parameters of Fig. \ref{fig:BulkTexture}. Bands are labeled by their Chern numbers (yellow squares). The green rectangle highlights the first bulk magnon gap.}
\label{fig:BulkBandTopo}
\end{figure}

%%%%%%%%%%%%%%%%%%%%%%%%%%%%%%%%%%%%%%%%%%%%%%%%%%%%%%%%%%%%%%%%%%%%%%
%%RESULTS
%%%%%%%%%%%%%%%%%%%%%%%%%%%%%%%%%%%%%%%%%%%%%%%%%%%%%%%%%%%%%%%%%%%%%%

{\it Topological Magnons and Edge States.} We first discuss the bulk magnon band structure. Figure \ref{fig:BulkBandTopo} shows the lowest-energy bands and their Chern numbers. No band vanishes at the gamma point (the lowest-energy band also has a tiny gap), i.e., there are no Goldstone modes \footnote{The absence of Goldstone modes was expected since the AFM-SkX is described by a discrete spin lattice configuration which exhibits neither translational nor rotational continuous symmetry.}. It should also be noted that the minimum of the second band is at a higher energy than the maximum of the first band, meaning that the first gap is global. Even though higher order gaps in Fig. \ref{fig:BulkBandTopo} can be found for each momentum, they are not global. Furthermore, the first band is topologically nontrivial: it has  Chern number $C_1 = -1$. By virtue of the bulk-edge correspondence \cite{Hatsugai1993a,Hatsugai1993b}, we expect chiral magnonic edge states with negative chirality within the first bulk magnon gap. These results are robust since they extend over a sizable region of the phase diagram. We find $C_1 = -1$ for all the AFM-SkX simulated values (blue and pink squares) from Fig. \ref{fig:PhDiag}.

\begin{figure}[t!]
\centering
\includegraphics[width=\columnwidth]{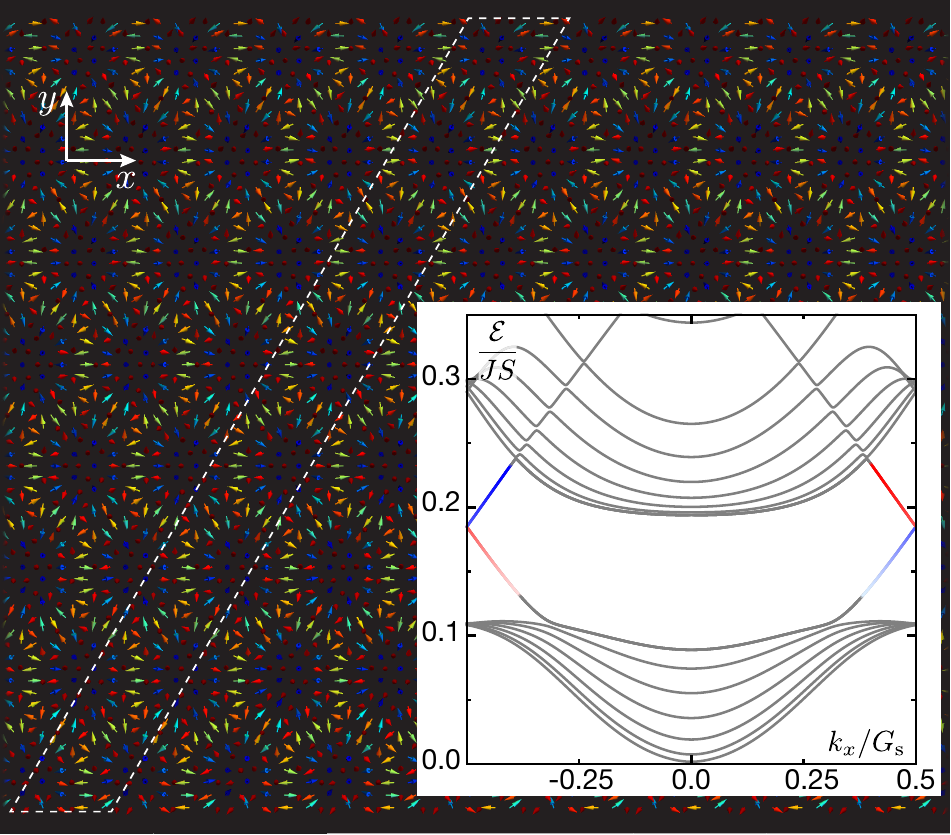}
\caption{AFM-SkX on a strip of infinite length along the $x$-axis with edges located at the top and bottom of the figure; parameters as in Fig. \ref{fig:BulkTexture}. The magnetic unit cell, comprised of 7$\times$63 spins, is enclosed by the white dashed-line loop. Inset: one-dimensional magnon band structure showing two chiral edge states within the first gap ($G_{\rm{s}} = 2\pi/7a$). The right-/left-moving edge state is depicted in blue/red while the color gradients encode the total magnon spin carried by each one.}
\label{fig:EdgeStateBand}
\end{figure}

\begin{figure}[t!]
\centering
\includegraphics[width=\columnwidth]{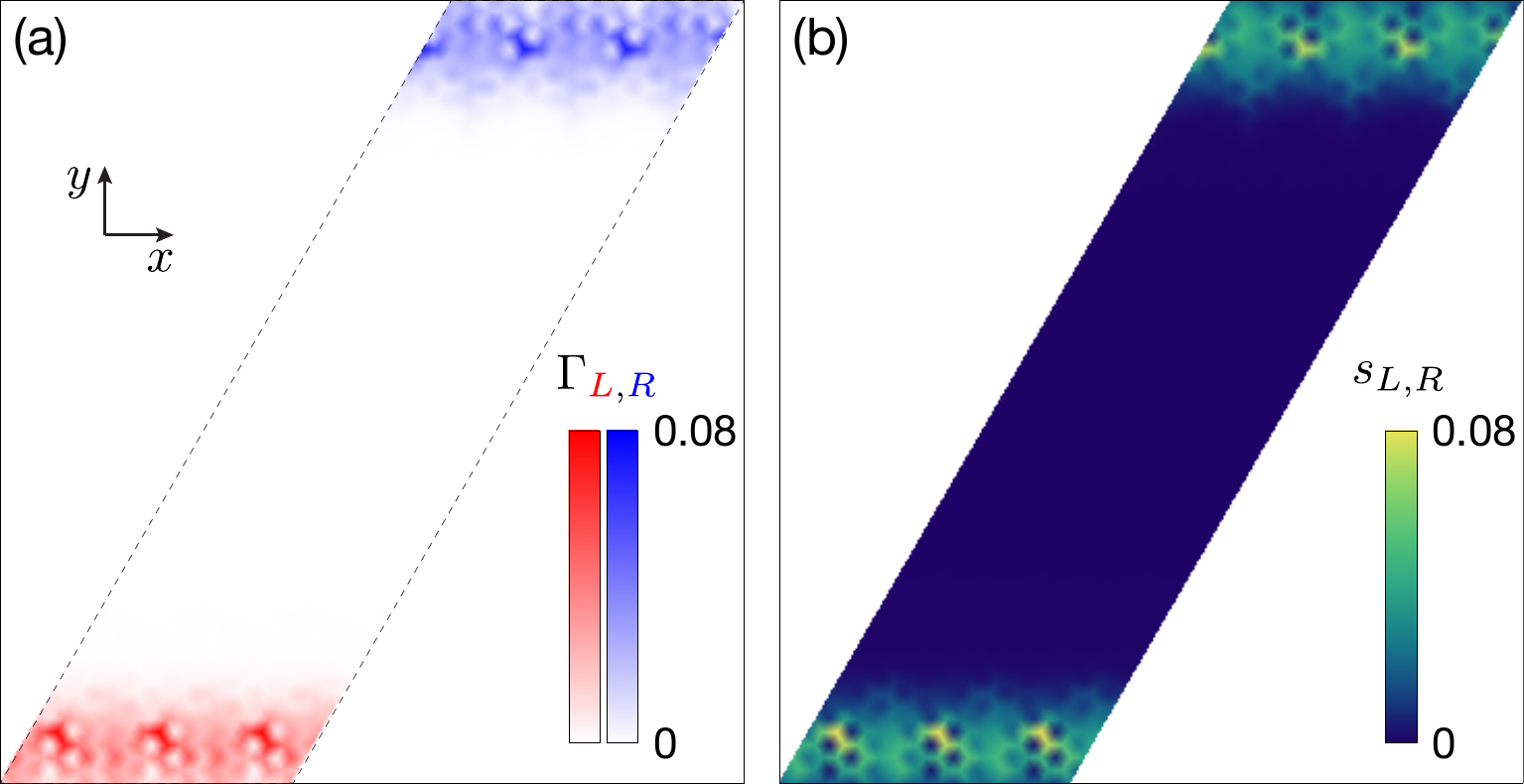}
\caption{Real space characterization of the chiral edge states from Fig. \ref{fig:EdgeStateBand} inset at $k_x = 0.45\,G_{\rm{s}}$. (a) Magnonic contribution of the wave function of the edge states, $\Gamma_{L,R}$, over an area spanning three times the magnetic unit cell. The right-/left-moving state, depicted in blue/red, is localized at the top/bottom, as expected of edge states with negative chirality. (b) Magnon spin density carried by the edge states.}
\label{fig:EdgeWF&Spin}
\end{figure}

For a strip geometry, however, the existence of the edge states needs to be proven explicitly since the bulk-edge correspondence relies on an integer number of unit cells within the strip \cite{Rhim2018}, which is generically not the case, see Fig. \ref{fig:EdgeStateBand}. Nevertheless, edge states emerge as we show next by allowing the texture to readjust to the new geometry. For this we recompute the classical ground state texture for open boundary conditions at the edges and periodic ones along the $x$-axis (see Fig. \ref{fig:EdgeStateBand}). After identifying the new magnetic unit cell, the one-dimensional magnon band structure is computed (see Fig. \ref{fig:EdgeStateBand} inset). Two states within the first gap have now appeared---connecting the bulk bands corresponding to the first and second bands from Fig. \ref{fig:BulkBandTopo}---that cross at $k_x = 0.5 G_{\rm{s}}$, with $G_{\rm{s}}$ the size of the one-dimensional BZ. Additionally, these in-gap states are chiral. We define the magnonic contribution of the right-/left-moving state wave function by $\Gamma_{R/L}(j,k_x) = |\bra{GS} a_j \, \alpha_{k_x R/L}^\dag \ket{GS}|^2$, where $j$ labels the site within the magnetic unit cell, see Fig. \ref{fig:EdgeWF&Spin}(a). For $k_x$ values corresponding to energies well within the global band gap, the right-/left-moving state is always localized at the top/bottom edge. On the other hand, when these states have an energy close to the bulk bands their wave functions are no longer edge-localized. Thus, these are indeed edge states \footnote{The two chiral edge states propagate on opposite edges of the infinite strip. In a finite-size system they would merge into a single chiral edge state with negative chirality, i.e., one that propagates clockwise.}.

Another quantity of interest is the magnon spin of these chiral edge states. The total magnon spin carried by the right-moving edge state is defined as $\BS_{R}(k_x) = \braket{k_x,R | \BS_{\rm{tot}} | k_x,R} - \braket{GS | \BS_{\rm{tot}} | GS}$, where $\BS_{\rm{tot}} = \sum_{\BR,j} \BS_{\BR j}$. To lowest order in the HP operator expansion series it becomes $\BS_{R}(k_x) = - \sum_j s_{R}(j,k_x) \, \Be_j^3$, with $s_{R}(j,k_x)$ the magnon spin density of the state $\ket{k_x,R} = \alpha_{k_x R}^\dag \ket{GS}$. Similar definitions are used for $\ket{k_x,L}$. In Fig. \ref{fig:EdgeStateBand} inset, $|\BS_{R}|$ and $|\BS_{L}|$ are color-coded in blue and red gradients (darker blue/red indicates higher magnon spin magnitude), and they both range between 0.73 and 0.83. Further, $s_{R}$ and $s_{L}$ are  localized at the top and bottom edge, respectively, as seen in Figure \ref{fig:EdgeWF&Spin}(b).

%%%%%%%%%%%%%%%%%%%%%%%%%%%%%%%%%%%%%%%%%%%%%%%%%%%%%%%%%%%%%%%%%%%%%%
%%DISCUSSION
%%%%%%%%%%%%%%%%%%%%%%%%%%%%%%%%%%%%%%%%%%%%%%%%%%%%%%%%%%%%%%%%%%%%%%

{\it Discussion.} Introducing edges in the strip geometry leads to a slightly distorted AFM-SkX from that of the infinite bulk case. For a sufficiently wide strip and well within its bulk, the distortions should decrease and the texture would resemble more the pristine AFM-SkX. Nevertheless, the fact that the chiral edge states we obtain are in full agreement with the bulk-edge correspondence, is further confirmation that continuous and small deformations in the Hamiltonian---in this case in the classical ground state texture---should affect neither the topology of the bulk bands nor the corresponding edge states as long as they do not close gaps. The same topological stability applies to the infinite bulk system. Changes in $D/J$ and magnetic field within the AFM-SkX region of the phase diagram from Fig. \ref{fig:PhDiag} do not close the first magnon gap and thus the Chern number of the first band remains at $C_1 = -1$.

Previous numerical studies of the magnonic band structure in ferromagnetic SkX's consistently found that the two lowest-energy bands were topologically trivial \cite{Roldan-Molina2016,Garst2017}. Our own numerical simulations also confirm these reports. Therefore, the lowest-energy edge states hosted by ferromagnetic SkX's can only appear at the third bulk magnon gap. On the other hand, the lowest-energy edge states in the AFM-SkX we study here lie within the first bulk magnon gap. This relevant distinction makes this AFM-SkX potentially more interesting for applications as small energies would be needed to excite magnonic edge states. More importantly, the higher the energy the more significant the magnon-magnon interactions become, resulting in linewidth broadening of high-energy magnon bands as confirmed by numerical calculations \cite{Mook2016b}. Therefore, the fact that the edge states emerge at low energies puts on a firmer footing the use of the noninteracting magnon approximation.

Skyrmions in two-sublattice antiferromagnets are expected to experience no skyrmion Hall effect due to their vanishing topological charge, an attractive property for applications that require skyrmions to follow a straight trajectory. In fact, a suppression of the skyrmion Hall effect was measured in two-sublattice ferrimagnetic thin films \cite{Woo2018}. In contrast, the three-sublattice structure of the AFM-SkX we study here is responsible for the $Q = -1$ topological charge carried by its magnetic unit cell. This nontrivial magnetic texture topology would therefore lead to the observation of the topological and skyrmion Hall effects in metallic systems. 

Topological magnons in a two-sublattice ferrimagnetic SkX have been studied analytically using a continuum model \cite{Kim2018}. At the compensation point, when the model effectively described an AFM-SkX, the edge states were found to be helical. It is possible that the external magnetic field that stabilizes the three-sublattice AFM-SkX or the emergent magnetic field due to its texture---fully incorporated in our description, while for simplicity taken as uniform in Ref. \cite{Kim2018}---favors the chiral edge states we find.
 
There are many materials reported in the literature as good candidate systems that realize a two-dimensional triangular Heisenberg antiferromagnet \cite{Collins1997,Nakatsuji2010}. Among them are the insulators NiGa$_2$S$_4$, NaCrO$_2$, and only a small group of them present DMI, such as CsCuCl$_3$. However, it appears possible to induce interfacial DMI in these systems by coupling them to heavy atoms in multilayers as has been done with materials supporting ferromagnetic skyrmions \cite{Soumyanarayanan2017}. The experimental detection of the magnonic edge states could be achieved combining state-of-the-art ferromagnetic resonance (FMR) \cite{Brataas2002,Du2017} and optical imaging \cite{Demokritov2001,Kruglyak2010} techniques. Magnons excited by FMR frequencies tuned to the energy window of the first bulk magnon gap would be observed only at the edges and not in the bulk of the sample. The fact that the edge states predicted here emerge at the lowest possible energies should also promise long magnonic lifetimes.

%%%%%%%%%%%%%%%%%%%%%%%%%%%%%%%%%%%%%%%%%%%%%%%%%%%%%%%%%%%%%%%%%%%%%
%%CONCLUSION
%%%%%%%%%%%%%%%%%%%%%%%%%%%%%%%%%%%%%%%%%%%%%%%%%%%%%%%%%%%%%%%%%%%%%%

{\it Conclusions.} We have shown that AFM-SkX's have topological magnon bands and moreover host topologically-protected chiral magnonic edge states in generic strip geometries. Furthermore, the lowest-energy edge states can emerge even within the first bulk magnon gap. These findings highlight the richness of the magnonic excitations in electrically insulating AFM-SkX's and their potential as platforms for topological magnonics where the properties that make antiferromagnets attractive for applications could be harnessed.

%%%%%%%%%%%%%%%%%%%%%%%%%%%%%%%%%%%%%%%%%%%%%%%%%%%%%%%%%%%%%%%%%%%%%%
\begin{acknowledgments}

We are grateful to Victor Chua for useful discussions. This work was supported by the Swiss National Science Foundation and  NCCR QSIT. This project received funding from the European Union's Horizon 2020 research and innovation program (ERC Starting Grant, grant agreement No 757725).

\end{acknowledgments}
%%%%%%%%%%%%%%%%%%%%%%%%%%%%%%%%%%%%%%%%%%%%%%%%%%%%%%%%%%%%%%%%%%%%%%

%
	
\end{document}